\documentclass[conference]{IEEEtran}
\IEEEoverridecommandlockouts
\IEEEpubid{
	\begin{minipage}{\textwidth}\ \\[12pt]
		\copyright 2022 IEEE\\ 
		DOI 10.1109/ICECS202256217.2022.9970992
\end{minipage}}

\usepackage{cite}
\usepackage{amsmath,amssymb,amsfonts}
\usepackage{graphicx}
\usepackage{textcomp}
\usepackage{xcolor}
\usepackage{multirow}
\usepackage{caption}
\usepackage{tabularx}
\usepackage{subfig}
\usepackage{mathtools}
\usepackage{color}
\usepackage{multicol}
\usepackage{flushend}
\usepackage{listings} 
\usepackage{algorithm}
\usepackage[noend]{algpseudocode}

\def\BibTeX{{\rm B\kern-.05em{\sc i\kern-.025em b}\kern-.08em
    T\kern-.1667em\lower.7ex\hbox{E}\kern-.125emX}}
\begin{document}

\title{Hardware-Software Co-Design of BIKE with HLS-Generated Accelerators\\
\thanks{This work was partially supported by SIAE MICROELETTRONICA and by
	the EU Horizon 2020 “TEXTAROSSA” project~(Grant No. 956831).}
}

\author{
	\IEEEauthorblockN{Gabriele Montanaro}
	\IEEEauthorblockA{\textit{DEIB}\\
		\textit{Politecnico di Milano}\\
		Milano, Italy\\
		gabriele.montanaro@polimi.it}
	\and
	\IEEEauthorblockN{Andrea Galimberti}
	\IEEEauthorblockA{\textit{DEIB}\\
		\textit{Politecnico di Milano}\\
		Milano, Italy\\
		andrea.galimberti@polimi.it}
	\and
	\IEEEauthorblockN{Ernesto Colizzi}
	\IEEEauthorblockA{\textit{SIAE MICROELETTRONICA}\\
		Milano, Italy\\
		ernesto.colizzi@siaemic.com}
	\and
	\IEEEauthorblockN{Davide Zoni}
	\IEEEauthorblockA{\textit{DEIB}\\
		\textit{Politecnico di Milano}\\
		Milano, Italy\\
		davide.zoni@polimi.it}
}

\maketitle

\IEEEpubidadjcol

\begin{abstract}
In order to mitigate the security threat of quantum computers,
NIST is undertaking a process to standardize post-quantum cryptosystems,
aiming to assess their security and speed up their adoption
in production scenarios.
Several hardware and software implementations have been proposed
for each candidate, while only a few target heterogeneous platforms
featuring CPUs and FPGAs.
This work presents a HW/SW co-design of BIKE for embedded platforms
featuring both CPUs and small FPGAs and employs high-level synthesis~(HLS)
to timely deliver the hardware accelerators.
In contrast to state-of-the-art solutions targeting performance-optimized
HLS accelerators, the proposed solution targets the small FPGAs implemented
in the heterogeneous platforms for embedded systems.
Compared to the software-only execution of BIKE, the experimental results
collected on the systems-on-chip of the entire Xilinx Zynq-7000 family
highlight a performance speedup ranging from 1.37$\times$, on Z-7010,
to 2.78$\times$, on Z-7020.
\end{abstract}


\section{Introduction and related works}
\label{sec:introduction}
In the near future, large-scale quantum computers are expected to break widely
used public-key cryptosystems, whose security relies on the hardness of
factoring large integers and computing discrete logarithms in a cyclic group.
To this end, post-quantum cryptography~(PQC) aims to design cryptoschemes
that can be executed on traditional, i.e., non-quantum, computers and
are secure against both traditional and quantum attacks.

In this scenario, the National Institute of Standards and Technology~(NIST)
undertook the process of evaluating and standardizing novel post-quantum schemes
to face the security threat imposed by the advances in quantum computing.
Given the wide range of scenarios that mandate the use of cryptographic primitives,
a goal of NIST is to ensure the possibility of implementing the selected
post-quantum cryptosystems on the largest variety of computing platforms.
Thus, efficient software and hardware implementations targeting
Intel Haswell CPUs and Xilinx Artix-7 FPGAs, respectively,
are critical factors in evaluating the NIST post-quantum candidates.
However, the actual adoption of PQC into production environments is subject to
the time-consuming process of designing and evaluating effective software and
hardware implementations of the candidate cryptosystems. To this end, the usage
of high-level synthesis~(HLS) emerged as a viable solution for timely delivery
of hardware implementations of PQC solutions~\cite{Dang_ePrint2020}.


Starting from the cryptosystems selected for the fourth evaluation round of the
NIST PQC contest~\cite{NIST_IR8413}, this work targets the hardware-software~(HW/SW) co-design
of the BIKE post-quantum key encapsulation module~(KEM), a candidate for future standardization
that is based on QC-MDPC codes~\cite{BIKE_website}.
The proposed HW/SW co-design of BIKE targets embedded platforms featuring both
CPUs and small FPGAs and employs HLS to design the hardware accelerators. 

HLS has been extensively used to deliver hardware implementations of
the candidates of the NIST PQC contest, including lattice-based KEMs~\cite{Dang_ePrint2020},
the Classic McEliece code-based KEM~\cite{Kostalabros_FPL2021}, and
comprehensive implementations of both lattice-based KEM and
digital signature schemes~\cite{Guerrieri_TechRxiv2022}.
A HW/SW co-design approach exploiting HLS to design hardware accelerators was
successfully employed targeting Classic McEliece~\cite{Kostalabros_FPL2021} and
lattice-based cryptosystems~\cite{Nguyen2020}.
Notably, the state-of-the-art contains few
hardware~\cite{Richter-Brockmann_TC2021,Richter-Brockmann_TCHES2022,
Barenghi_ICECS2019, Zoni_Access2020_Mul, Zoni_Access2020_Dec,
Galimberti_TC2022}
and software~\cite{BIKE_website, Drucker_CSCML2020,Chen_ACNS2022}
BIKE implementations, while, to the best of our knowledge, no HW/SW
co-design solution was proposed. 

\smallskip\noindent\textbf{Contributions} -
In contrast to existing state-of-the-art
solutions targeting performance-optimized HLS accelerators, the proposed HW/SW
co-design approach aims to optimize the area-performance trade-off for those embedded
computing platforms featuring both a CPU and programmable logic. 
Notably, optimizing performance is subject to the limited programmable
hardware resources of the considered platforms and thus represents
an additional and challenging design factor when using HLS to design
the hardware accelerators.

Compared to the reference software execution of BIKE, the results of the
proposed HW/SW co-design targeting the Xilinx Zynq-7000 embedded-class SoC
family, i.e., Z-7010, Z-7015, and Z-7020, show performance improvements up to 2.78x.

\begin{figure*}[!h]
	\scalebox{0.98}{
		\begin{minipage}[t]{0.315\linewidth}
			\begin{algorithm}[H]
				\caption{Key generation.}
				\label{alg:key_generation}
				\begin{algorithmic}[1]
					\Function{$[H, \sigma, h]$ KeyGen\ }{\ }
					\State $seed = $ TRNG $();$
					\State $H = $ PRNG$($SHAKE$(seed));$
					
					\State $f = h_0; res = h_0;$
					\For{$i \in 1 $ to $\lfloor log_2 (p-2) \rfloor$}
						\State$f = f \odot f^{2^{2^{i-1}}};$
						\If{$(p-2)_2[i] = 1_2$}
							\State$res = res \odot f^{2^{r-2 \bmod 2^i}};$
						\EndIf
					\EndFor
					\State$h_{0_{inv}} = f^2;$
					
					\State $h = h_1 \odot h_{0_{inv}};$
					\State $\sigma = $ TRNG $();$
					\State\Return $\{H, \sigma, h\};$
					\EndFunction
				\end{algorithmic}
			\end{algorithm}
		\end{minipage}%
	}
	\hfill
	\scalebox{0.98}{
		\begin{minipage}[t]{0.265\linewidth}
			\begin{algorithm}[H]
				\caption{Encapsulation.}
				\label{alg:encapsulation}
				\begin{algorithmic}[1]
					\Function{$[K, c]$ Encaps\ }{$h$}
					\State $m = $ TRNG $();$
					\State $e = $ PRNG$($SHAKE$(m));$
					\State $s = e_0 \oplus (e_1 \odot h);$
					\State $m' = m \oplus $ SHA3$(e));$ 
					\State $c = \{s, m'\};$
					\State $K = $SHA3$(\{m,c\}));$
					\State\Return $\{K, c\};$
					\EndFunction
				\end{algorithmic}
			\end{algorithm}
		\end{minipage}%
	}
	\hfill
	\scalebox{0.98}{
		\begin{minipage}[t]{0.40\linewidth}
			\begin{algorithm}[H]
				\caption{Decapsulation.}
				\label{alg:decapsulation}
				\begin{algorithmic}[1]
					\Function{$[K]$ Decaps\ }{$H$, $\sigma$, $c$}
					\State $s' = h_0 \odot s;$

					\State$e' = 0;$
					\While{$s' \neq 0$}
						\State$upc = s' \cdot H;$
						\State$e' = e' \oplus (upc \geq thr);$
						\State$s' = e' \odot H^T;$
					\EndWhile
					
					\State $m'' = m' \oplus $ SHA3$(e'));$
					\State $a = e' = $ PRNG$($SHAKE$(m''))$ ? $m''$ : $\sigma;$
					\State$K = $ SHA3$(\{a, c\}));$
					\State\Return $K;$
					\EndFunction
				\end{algorithmic}
			\end{algorithm}
		\end{minipage}
	}
	\caption{Algorithms for the key generation, encapsulation, and decapsulation primitives of BIKE~\cite{BIKE_website}.}
	\label{fig:alg_bike}
\end{figure*}

\section{Methodology}
\label{sec:methodology}
\subsection{BIKE specification and baseline HLS}
\label{sec:methodology_originalSW}
%
%
%
Figure~\ref{fig:alg_bike} shows the algorithms for the three main primitives of
BIKE, i.e., key generation~(Algorithm~\ref{alg:key_generation}),
encapsulation~(Algorithm~\ref{alg:encapsulation}), and
decapsulation~(Algorithm~\ref{alg:decapsulation}).
Notably, few critical operations dominate the computational
complexity, thus representing the leading candidates for optimization
in the HLS process.
The key generation requires a binary polynomial inversion~(see lines 4-9 in
Algorithm~\ref{alg:key_generation}), a binary polynomial multiplication~(line
10), and SHAKE256-based sampling~(line 3).
The encapsulation requires a binary polynomial multiplication~(see line 4 in
Algorithm~\ref{alg:encapsulation}), uniform random sampling employing
SHAKE256~(line 3), and the computation of two SHA3-384 hash digests~(lines 5
and 7).
The decapsulation requires a binary polynomial multiplication~(see line 2 in
Algorithm~\ref{alg:decapsulation}), QC-MDPC bit-flipping decoding~(lines 3-7),
computing SHA3-384 digests~(lines 8 and 10), and SHAKE256-based sampling~(line
9).
%

%
\smallskip
\noindent\textbf{Baseline HLS implementation -} Preliminary changes to the
original software are mandatory to meet the HLS specification requirements.
Unbounded arrays passed as arguments by pointer are replaced with bounded
arrays. Moreover, the original recursive formulation of the multiplication is
not supported by the HLS frameworks, therefore it was replaced with
a simpler Comba implementation. 

\subsection{HLS optimizations and HW/SW co-design}
\label{sec:methodology_optimizations}
The proposed co-design approach is organized in three steps to deliver an
area-performance optimized HW/SW solution. 
The \emph{performance optimization} step aims to optimize the
execution time of each of the three primitives of BIKE separately. The
subsequent \emph{area optimization} step targets the resource utilization of
each performance-optimized primitive of BIKE.
Last, the \emph{HW/SW co-design} step delivers the final solution
by selectively implementing each primitive either in hardware or software to
maximize the area-performance trade-off. 

\smallskip\noindent\textbf{Performance optimization -} 
Starting from the baseline designs, we explored the most time-consuming
operations of each primitive.
In particular, multiplication is a critical operation in all KEM
primitives while also dominating the execution time for both
the key generation and the decapsulation primitives.
%
%
%
We rewrite the multiplication code to speed up all three KEM primitives
by adding a Karatsuba multiplication layer~\cite{Karatsuba_1962} on top of
the Comba multiplication~\cite{Comba_IBM1990}.
Notably, the proposed design allows configuring the number of Karatsuba recursions
at compile-time to allow a configurable area-performance trade-off.
In addition, applying \emph{loop unrolling} and \emph{loop pipelining}
to the innermost Comba multiplication logic significantly reduces
the latency of multiplications.

\begin{table*}[t]
	\centering
	\caption{Comparison between software and HLS-based implementations across high-level synthesis optimization process.
	}
	\scalebox{1}{
		\begin{tabular}{cccccccccc}
			\multirow{2}{*}{\textbf{Target}} &            \textbf{Design}             &    \textbf{KEM}    &    LUT    &    FF     &    DSP    &   BRAM    & Clock &     Latency      &  Speedup   \\
			                                 &        \textbf{(Optimization)}         & \textbf{Primitive} & [\#~(\%)] & [\#~(\%)] & [\#~(\%)] & [\#~(\%)] & [MHz] & [ms ($10^3$ cc)] & [$\times$] \\ \hline\hline
			 \multirow{3}{*}{\textbf{CPU}}   & \multirow{3}{*}{\textbf{Baseline SW}}  &  \textbf{KeyGen}   &    $-$    &    $-$    &    $-$    &    $-$    &  667  & \ 332.14~(221537)  &     1      \\
			                                 &                                        &  \textbf{Encaps}   &    $-$    &    $-$    &    $-$    &    $-$    &  667  &   14.86~(9913)   &     1      \\
			                                 &                                        &  \textbf{Decaps}   &    $-$    &    $-$    &    $-$    &    $-$    &  667  & \ 464.61~(309894)  &     1      \\ \hline
			 \multirow{9}{*}{\textbf{FPGA}}  & \multirow{3}{*}{\textbf{Baseline HLS}} &  \textbf{KeyGen}   &   14110   &   10011   &     0     &    28     &  100  &  268.67~(26867)  &    1.24    \\
			                                 &                                        &  \textbf{Encaps}   &   64097   &   56581   &     0     &    93     &  100  &   16.69~(1669)   &    0.89    \\
			                                 &                                        &  \textbf{Decaps}   &  106799   &   86432   &     0     &    169    &  100  &  248.96~(24896)  &    1.86    \\ \cline{2-10}
			                                 & \multirow{2}{*}{\textbf{Interm. HLS}}  &  \textbf{KeyGen}   &   17208   &   14428   &     0     &    36     &  100  &  137.83~(13783)  &    2.41    \\
			                                 &    \multirow{2}{*}{\textbf{(Perf)}}    &  \textbf{Encaps}   &   66887   &   59219   &     0     &    129    &  100  &    6.49~(649)    &    2.29    \\
			                                 &                                        &  \textbf{Decaps}   &  120918   &   95953   &    14     &    193    &  100  &  135.70~(13570)  &    3.42    \\ \cline{2-10}
			                                 &  \multirow{2}{*}{\textbf{Final HLS}}   &  \textbf{KeyGen}   &   13567   &   11621   &     0     &    40     &  100  &  137.84~(13784)  &    2.41    \\
			                                 & \multirow{2}{*}{\textbf{(Perf+Area)}}  &  \textbf{Encaps}   &   23260   &   15571   &     0     &    96     &  100  &    6.33~(633)    &    2.35    \\
			                                 &                                        &  \textbf{Decaps}   &   37160   &   38118   &    35     &    90     &  100  &  135.48~(13548)  &    3.43    \\ \hline
		\end{tabular}}
	\label{tbl:hls_results}
\end{table*}

\noindent\textbf{Area optimization} -
Area optimization is carried out first by enforcing resource sharing,
employing the \emph{function inlining} and \emph{resource allocation} HLS
directives.
%
%
%
Resource sharing was enforced within the bit-flipping decoding,
multiplication, SHA-3, and SHAKE operations.
In particular, we instantiate the common logic of SHA-3 and SHAKE
only once within each KEM primitive since the two share a
significant amount of C code, drastically reducing their
occupied area.
Since multiplication also appears in key generation while
encapsulation employs multiplication, SHA-3, and SHAKE,
the area of all three KEM primitives is actually reduced
by applying the aforementioned changes.
In addition, struct variables, which used a multitude of LUT resources,
were modified into array variables, saving a significant amount of area.
Moreover, the \emph{storage binding} HLS directive was used to force
the implementation of small variables as RAM instead of ROM memories,
for which the default implementation consumed too many BRAM blocks.
Last, \emph{array partitioning} directives were employed to reduce BRAM
utilization, which otherwise would end up as the scarcest resource due to
the many array variables declared in the C code.
Such optimization allowed indeed to force the usage of flip-flops, instead of BRAM,
for the smaller variables, such as 32-bit $seed$ and 256-bit $\sigma$, $m$, $m'$, and $m''$
detailed in Figure~\ref{fig:alg_bike}.
Notably, due to the large size of the polynomials, in the order of thousands of bits,
variables holding polynomial data are instead left mapped to BRAM.

\smallskip\noindent\textbf{HW/SW co-design -}
The HW/SW co-design phase aims to identify the best mix of
KEM primitives executed on the CPU and instantiated on the FPGA,
depending on the performance of the software execution and the HLS modules
subject to the resource utilization of the latter.
The identified solution must minimize latency while satisfying
the area constraints given by the FPGA part of the target SoC.
The exploration of the possible HW/SW combinations will prioritize
hardware modules that provide the most significant latency reductions and
that occupy the smallest amount of FPGA resources.
%

\section{Experimental evaluation}
\label{sec:expEval}
This section discusses the results of the HW-SW co-design of BIKE with NIST
security level 1, i.e., security against quantum attacks equivalent to AES-128,
targeting the Z-7010, Z-7015, and Z-7020 Xilinx Zynq-7000 SoCs.

\subsection{Experimental setup}
\label{ssec:expEval_setup}
%
The reference software execution was carried out on the CPU
part of the Xilinx Zynq-7000 SoC,
executing the Xilinx Petalinux 2022.1 operating system.
The Zynq-7000 chips feature a 32-bit dual-core ARM Cortex-A9 processor
that implements the ARM v7 ISA and runs at a 667MHz clock frequency.
The software execution targeted the C99 reference implementation
of BIKE~\cite{BIKE_website}.

The high-level synthesis of the hardware components was carried
out through Xilinx Vitis HLS 2022.1, starting from  the portable
optimized C implementation of BIKE~\cite{BIKE_SW_Github}.
The high-level synthesis and the RTL synthesis and implementation
via Xilinx Vivado 2022.1 targeted the FPGA parts of the
Xilinx Zynq-7000 Z-7010, Z-7015, and Z-7020
chips, feeding them a 100MHz clock frequency.
%
%
The available FPGA resources consist of
17600, 46200, and 53200 look-up tables~(LUT),
35200, 92400, and 106400 flip-flops~(FF),
80, 160, and 220 DSP slices~(DSP), and
60, 95, and 140 36Kb blocks of block RAM~(BRAM), respectively.
The area results reported in the following
were obtained after RTL implementation.



\begin{table}[t]
	\centering
	\caption{Area and performance comparison between software, hardware, and hardware-software solutions.
		The $-$ mark denotes no resources used due to software execution.
	}
	\scalebox{1}{
		\begin{tabular}{ccccccc}
			\multirow{2}{*}{\textbf{Design}} &    \textbf{KEM}    &  LUT  &   FF   & DSP  & BRAM & Latency \\
			                                 & \textbf{primitive} & [\#]  &  [\#]  & [\#] & [\#] &  [ms]   \\ \hline\hline
			  \multirow{4}{*}{\textbf{SW}}   &  \textbf{KeyGen}   &  $-$  &  $-$   & $-$  & $-$  & 332.14  \\
			                                 &  \textbf{Encaps}   &  $-$  &  $-$   & $-$  & $-$  &  14.86  \\
			                                 &  \textbf{Decaps}   &  $-$  &  $-$   & $-$  & $-$  & 464.61  \\ \cline{2-7}
			                                 &   \textbf{Total}   &  $-$  &  $-$   & $-$  & $-$  & 811.61  \\ \hline
			\multirow{4}{*}{\textbf{Z-7010}} &  \textbf{KeyGen}   & 13567 & 11621  &  0   &  40  & 137.84  \\
			\multirow{4}{*}{\textbf{HW/SW}}  &  \textbf{Encaps}   &  $-$  &  $-$   & $-$  & $-$  &  14.86  \\
			                                 &  \textbf{Decaps}   &  $-$  &  $-$   & $-$  & $-$  & 464.61  \\ \cline{2-7}
			                                 &   \textbf{Total}   & 13567 & 11621  &  0   &  40  & 617.31  \\ \cline{2-6}
			                                 & \textbf{Available} & 17600 & 35200  &  80  &  60  &         \\ \hline
			\multirow{4}{*}{\textbf{Z-7015}} &  \textbf{KeyGen}   &  $-$  &  $-$   & $-$  & $-$  & 332.14  \\
			\multirow{4}{*}{\textbf{HW/SW}}  &  \textbf{Encaps}   &  $-$  &  $-$   & $-$  & $-$  &  14.86  \\
			                                 &  \textbf{Decaps}   & 37160 & 38118  &  35  &  90  & 135.48  \\ \cline{2-7}
			                                 &   \textbf{Total}   & 37160 & 38118  &  35  &  90  & 482.48  \\ \cline{2-6}
			                                 & \textbf{Available} & 46200 & 92400  & 160  &  95  &         \\ \hline
			\multirow{4}{*}{\textbf{Z-7020}} &  \textbf{KeyGen}   & 13567 & 11621  &  0   &  40  & 137.84  \\
			\multirow{4}{*}{\textbf{HW/SW}}  &  \textbf{Encaps}   &  $-$  &  $-$   & $-$  & $-$  &  14.86  \\
			                                 &  \textbf{Decaps}   & 37160 & 38118  &  35  &  90  & 135.48  \\ \cline{2-7}
			                                 &   \textbf{Total}   & 50727 & 49739  &  35  & 130  & 288.18  \\ \cline{2-6}
			                                 & \textbf{Available} & 53200 & 106400 & 220  & 140  &         \\ \hline
			  \multirow{4}{*}{\textbf{HW}}   &  \textbf{KeyGen}   & 13567 & 11621  &  0   &  40  & 137.84  \\
			                                 &  \textbf{Encaps}   & 23260 & 15571  &  0   &  96  &  6.33   \\
			                                 &  \textbf{Decaps}   & 37160 & 38118  &  35  &  90  & 135.48  \\ \cline{2-7}
			                                 &   \textbf{Total}   & 73987 & 65310  &  35  & 226  & 279.65  \\ \hline
		\end{tabular}}
	\label{tbl:hwsw_results}
\end{table}

\subsection{Experimental results}
\label{ssec:expEval_results}
Table~\ref{tbl:hls_results} details the resource utilization of
the HLS-based implementations of BIKE and compares
their performance with the reference software execution. 
Resource utilization is expressed as the absolute amount of look-up tables~(LUT),
flip-flops~(FF), DSP slices~(DSP), and 36Kb blocks of block RAM~(BRAM) and
their relative utilization of the resources available on the target FPGA.
Performance statistics are reported in terms of the clock frequency, expressed in MHz,
and the latency, expressed in milliseconds and thousands of clock cycles.
In addition, the speedup metric represents the ratio between
the execution time of the reference software execution and
the latency of the current target.

\smallskip\noindent\textbf{High-level synthesis optimization} -
In this paragraph, we discuss the improvements to the KEM primitive
modules across the HLS optimization process, referring to
the experimental results detailed in Table~\ref{tbl:hls_results}.

Compared to the software execution of BIKE, the \emph{Baseline HLS} designs report
a performance speedup of 1.24$\times$ and 1.86$\times$ for key generation and decapsulation,
respectively, while the encapsulation primitive was slightly slower.
The three HLS modules occupy a large number of resources, particularly
LUT and BRAM ones, with only the \emph{KeyGen} one fitting in
the Zynq-7000 chips.

After performance optimization, the \emph{Interm. HLS} designs
are at least 2$\times$ faster than software execution, with
a speedup up to 3.42$\times$ for decapsulation, at the cost of increased area.

At the end of the area optimization step, the \emph{Final HLS} designs obtained
from the area optimization step exhibit a large resource utilization
reduction with negligible performance penalties.
The \emph{Decaps} module fits even in the intermediate Zynq-7000 SoC,
i.e., Z-7015, while the combined \emph{KeyGen} and
\emph{Decaps} modules can be concurrently implemented on Z-7020.
Notably, the area-optimized \emph{Decaps} module saves more than 80000 LUTs, 57000
FFs, and 100 BRAMs compared to the baseline HLS design.

\begin{figure}[t]
	\centering
	\includegraphics[width=\columnwidth]{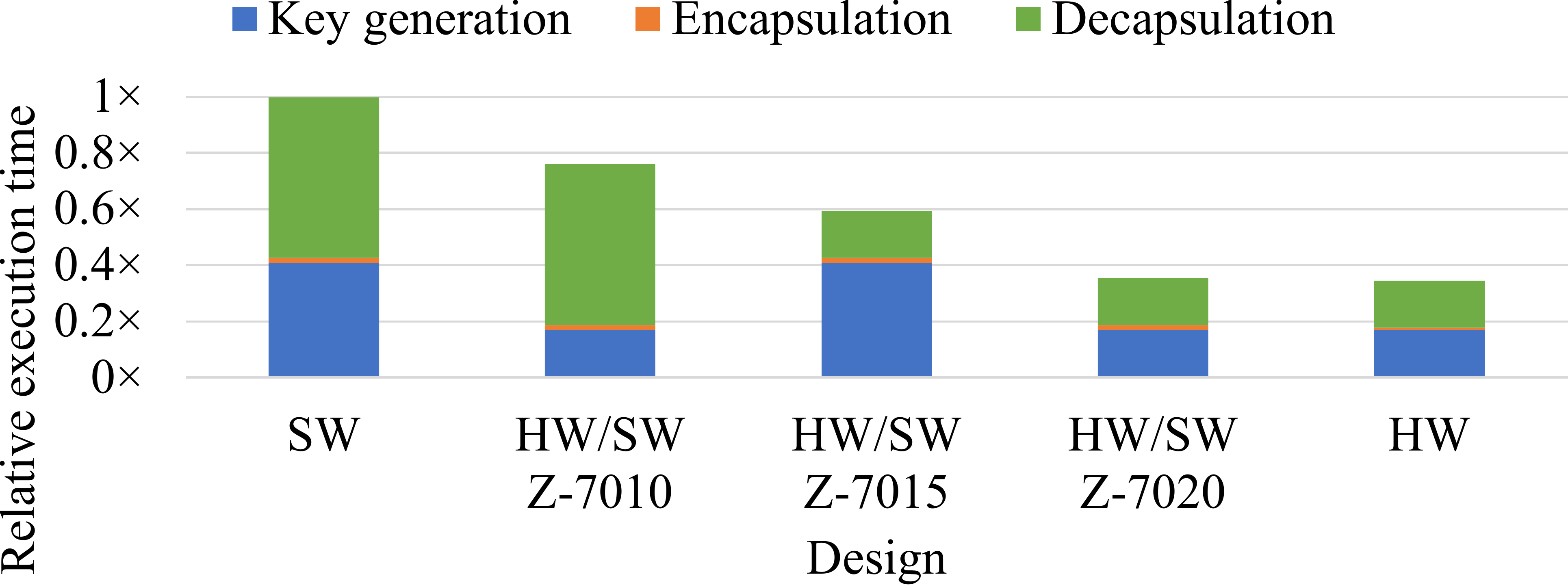}
	\caption{Relative execution time, normalized to reference software execution (lower is better).}
	\label{fig:rel_exec_time}
\end{figure}

\smallskip\noindent\textbf{Hardware-software co-design} -
Table~\ref{tbl:hwsw_results} details the resource utilization
and performance of the identified HW/SW solutions, comparing
them to the reference software execution and the hardware
instantiation of all three KEM primitives.
In addition, the execution time, normalized to reference software execution,
is represented in Figure~\ref{fig:rel_exec_time}.
The KEM primitives implemented in hardware are chosen to minimize
latency while fitting into each of the three target Zynq-7000 chips.

The HW-SW co-design solution targeting the Z-7010 SoC delivers a 1.31$\times$
performance speedup, i.e., 0.76$\times$ the latency of software-only execution,
implementing the \emph{KeyGen} module in hardware while the other two KEM primitives are
executed in software.

%
The identified Z-7015 design provides a 1.70x performance
speedup, i.e., 0.59$\times$ the latency of software-only execution, implementing
the \emph{Decaps} module in hardware while the other two KEM primitives are
executed in software.

%

Finally, applying our HW/SW co-design approach to the larger Z-7020 chip
results in a 2.78x performance speedup, i.e., 0.36$\times$ the latency of software-only execution,
implementing both \emph{KeyGen} and \emph{Decaps} modules in hardware while
\emph{Encaps} is still executed in software.

\section{Conclusions}
\label{sec:conclusions}
This work presents an HW/SW co-design of BIKE for those
embedded platforms featuring both CPUs and small FPGAs and
employs high-level synthesis~(HLS) to timely deliver the
hardware accelerators.
In contrast to state-of-the-art solutions targeting
performance-optimized HLS accelerators, the proposed
solution offers an area-performance optimized co-design
targeting the small FPGAs implemented in heterogenous
embedded platforms.
Compared to the software-only execution of BIKE, the
experimental results collected on the systems-on-chip of
the entire Xilinx Zynq-7000 family highlight a performance
speedup ranging from 1.37$\times$, on Z-7010, to
2.78$\times$, on Z-7020.


\bibliographystyle{IEEEtran}
\bibliography{2022_ICECS}

\begin{thebibliography}{10}
\providecommand{\url}[1]{#1}
\csname url@samestyle\endcsname
\providecommand{\newblock}{\relax}
\providecommand{\bibinfo}[2]{#2}
\providecommand{\BIBentrySTDinterwordspacing}{\spaceskip=0pt\relax}
\providecommand{\BIBentryALTinterwordstretchfactor}{4}
\providecommand{\BIBentryALTinterwordspacing}{\spaceskip=\fontdimen2\font plus
\BIBentryALTinterwordstretchfactor\fontdimen3\font minus
  \fontdimen4\font\relax}
\providecommand{\BIBforeignlanguage}[2]{{%
\expandafter\ifx\csname l@#1\endcsname\relax
\typeout{** WARNING: IEEEtran.bst: No hyphenation pattern has been}%
\typeout{** loaded for the language `#1'. Using the pattern for}%
\typeout{** the default language instead.}%
\else
\language=\csname l@#1\endcsname
\fi
#2}}
\providecommand{\BIBdecl}{\relax}
\BIBdecl

\bibitem{Dang_ePrint2020}
\BIBentryALTinterwordspacing
V.~B. Dang, F.~Farahmand, M.~Andrzejczak, K.~Mohajerani, D.~T. Nguyen, and
  K.~Gaj, ``Implementation and benchmarking of round 2 candidates in the nist
  post-quantum cryptography standardization process using hardware and
  software/hardware co-design approaches,'' Cryptology ePrint Archive, Paper
  2020/795, 2020, \url{https://eprint.iacr.org/2020/795}. [Online]. Available:
  \url{https://eprint.iacr.org/2020/795}
\BIBentrySTDinterwordspacing

\bibitem{NIST_IR8413}
{National Institute of Standards and Technology (NIST) - U.S. Department of
  Commerce}, ``Nistir 8413, status report on the third round of the nist
  post-quantum cryptography standardization process,''
  \url{https://nvlpubs.nist.gov/nistpubs/ir/2022/NIST.IR.8413.pdf}, 2022.

\bibitem{BIKE_website}
N.~Aragon, P.~S. L.~M. Barreto, S.~Bettaieb, L.~Bidoux, O.~Blazy, J.-C.
  Deneuville, P.~Gaborit, S.~Gueron, T.~G\"{u}neysu, C.~A. Melchor,
  R.~Misoczki, E.~Persichetti, N.~Sendrier, J.-P. Tillich, V.~Vasseur, and
  G.~Z{\'e}mor, ``{BIKE} website,'' \url{https://www.bikesuite.org/}, 2021.

\bibitem{Kostalabros_FPL2021}
V.~Kostalabros, J.~Ribes-González, O.~Farràs, M.~Moretó, and C.~Hernandez,
  ``Hls-based hw/sw co-design of the post-quantum classic mceliece
  cryptosystem,'' in \emph{2021 31st International Conference on
  Field-Programmable Logic and Applications (FPL)}, 2021, pp. 52--59.

\bibitem{Guerrieri_TechRxiv2022}
\BIBentryALTinterwordspacing
A.~Guerrieri, G.~D.~S. Marques, F.~Regazzoni, and A.~Upegui, ``{Design
  Exploration and Code Optimizations for FPGA-Based Post-Quantum Cryptography
  using High-Level Synthesis},'' 3 2022. [Online]. Available:
  \url{https://doi.org/10.36227/techrxiv.19404413.v1}
\BIBentrySTDinterwordspacing

\bibitem{Nguyen2020}
D.~T. Nguyen, V.~B. Dang, and K.~Gaj, ``High-level synthesis in implementing
  and benchmarking number theoretic transform in lattice-based post-quantum
  cryptography using software/hardware codesign,'' in \emph{Applied
  Reconfigurable Computing. Architectures, Tools, and Applications},
  F.~Rinc{\'o}n, J.~Barba, H.~K.~H. So, P.~Diniz, and J.~Caba, Eds.\hskip 1em
  plus 0.5em minus 0.4em\relax Cham: Springer International Publishing, 2020,
  pp. 247--257.

\bibitem{Richter-Brockmann_TC2021}
J.~Richter-Brockmann, J.~Mono, and T.~Guneysu, ``Folding bike: Scalable
  hardware implementation for reconfigurable devices,'' \emph{IEEE Transactions
  on Computers}, 2021.

\bibitem{Richter-Brockmann_TCHES2022}
\BIBentryALTinterwordspacing
J.~Richter-Brockmann, M.-S. Chen, S.~Ghosh, and T.~Güneysu, ``Racing bike:
  Improved polynomial multiplication and inversion in hardware,'' Cryptology
  ePrint Archive, Paper 2021/1344, 2021,
  \url{https://eprint.iacr.org/2021/1344}. [Online]. Available:
  \url{https://eprint.iacr.org/2021/1344}
\BIBentrySTDinterwordspacing

\bibitem{Barenghi_ICECS2019}
A.~Barenghi, W.~Fornaciari, A.~Galimberti, G.~Pelosi, and D.~Zoni, ``Evaluating
  the trade-offs in the hardware design of the ledacrypt encryption
  functions,'' in \emph{2019 26th IEEE International Conference on Electronics,
  Circuits and Systems (ICECS)}, 2019, pp. 739--742.

\bibitem{Zoni_Access2020_Mul}
D.~{Zoni}, A.~{Galimberti}, and W.~{Fornaciari}, ``Flexible and scalable
  fpga-oriented design of multipliers for large binary polynomials,''
  \emph{IEEE Access}, vol.~8, pp. 75\,809--75\,821, 2020.

\bibitem{Zoni_Access2020_Dec}
------, ``Efficient and scalable fpga-oriented design of qc-ldpc bit-flipping
  decoders for post-quantum cryptography,'' \emph{IEEE Access}, vol.~8, pp.
  163\,419--163\,433, 2020.

\bibitem{Galimberti_TC2022}
A.~Galimberti, G.~Montanaro, and D.~Zoni, ``Efficient and scalable fpga design
  of gf(2m) inversion for post-quantum cryptosystems,'' \emph{IEEE Transactions
  on Computers}, pp. 1--1, 2022.

\bibitem{Drucker_CSCML2020}
N.~Drucker, S.~Gueron, and D.~Kostic, ``Fast polynomial inversion for post
  quantum qc-mdpc cryptography,'' in \emph{Cyber Security Cryptography and
  Machine Learning}, S.~Dolev, V.~Kolesnikov, S.~Lodha, and G.~Weiss,
  Eds.\hskip 1em plus 0.5em minus 0.4em\relax Cham: Springer International
  Publishing, 2020, pp. 110--127.

\bibitem{Chen_ACNS2022}
M.-S. Chen, T.~G{\"u}neysu, M.~Krausz, and J.~P. Thoma, ``Carry-less to bike
  faster,'' in \emph{Applied Cryptography and Network Security}, G.~Ateniese
  and D.~Venturi, Eds.\hskip 1em plus 0.5em minus 0.4em\relax Cham: Springer
  International Publishing, 2022, pp. 833--852.

\bibitem{Karatsuba_1962}
A.~{Karatsuba} and Y.~{Ofman}, ``Multiplication of many-digital numbers by
  automatic computers,'' \emph{Proceedings of the USSR Academy of Sciences},
  vol. 145, pp. 293--294, 1962.

\bibitem{Comba_IBM1990}
P.~G. {Comba}, ``Exponentiation cryptosystems on the ibm pc,'' \emph{IBM
  Systems Journal}, vol.~29, no.~4, pp. 526--538, 1990.

\bibitem{BIKE_SW_Github}
{Amazon Web Services - Labs}, ``Additional implementation of bike (bit flipping
  key encapsulation),'' \url{https://github.com/awslabs/bike-kem}, 2020.

\end{thebibliography}

\end{document}